\documentstyle[psfig,twocolumn,aps]{revtex}
\begin {document}
\draft
\title{Second bound state of the positronium molecule and biexcitons}
\author{K. Varga$^{1,2}$, J. Usukura$^3$ and Y. Suzuki$^4$}
\address{
$^1$Institute of Nuclear Research of the Hungarian Academy of Sciences 
(ATOMKI), Debrecen, H-4001, Hungary
\\
$^2$Institute for Physical and Chemical Research (RIKEN), Wako, Saitama
351-01, Japan
\\
$^3$Graduate School of Science and Technology, Niigata University,
Niigata 950--21, Japan
\\
$^4$Department of Physics, Faculty of Science, Niigata University,
Niigata 950--21, Japan}

\date{\today}

\maketitle

\begin{abstract}
A new, hitherto unknown
bound state of the positronium molecule, with orbital angular momentum
$L=1$ and negative parity is reported. This state is stable against 
autodissociation even if the masses of the positive and negative charges 
are not equal. The existence of a similar state in two-dimension has also
been investigated. The fact that the biexcitons have a second bound state
may help the better understanding of their binding mechanism.
\end{abstract}
\pacs{PACS numbers: 36.10.Dr, 31.15.Pf, 71.35.-y, 73.20.Dx}

\narrowtext
Although the hydrogen and positronium molecule are both 
quantum-mechanical
fermionic four-body systems of two positively and two negatively charged 
identical particles on the two opposite ends of the 
$(M+,M+,m-,m-)$-type Coulombic systems,
their properties are extremely different. Their experimental and 
theoretical ``history'' reflects these dissimilarities. The 
spectrum of 
the hydrogen molecule has already been known in the last century
long before the birth of quantum mechanics,
and the first theoretical calculation of Heitler and London is 70 
years old \cite{HL}. 
The existence of the Ps$_2$ molecule
was first theoretically predicted by Ore \cite{Ore} half a century ago. 
The positronium molecule (Ps$_2$), however, has not been
experimentally found yet, and therefore the theoretical calculations
are especially important. 
\par\indent
Between the H$_2$ and Ps$_2$ molecule
one can continuously change the mass ratio $\sigma=m/M$ and 
the particles 
form a bound molecule. In solid state physics 
these  ``biexcitons'' (system of 
two holes and two electrons) play a particularly important role.
The typical mass ratio in the biexcitonic molecules is $m/M=0.67$, therefore
the properties of these species are rather similar to that of  Ps$_2$. 
A few years ago the biexcitons were experimentally observed 
in GaAs/AlGaAs quantum wells \cite{Miller}.
\par\indent
While in the case of the  H$_2$ molecule many bound excited states
have been observed and later theoretically studied, in the case of
the Ps$_2$ molecule only the ground state has been predicted.
Unlike the H$_2$ molecule, the Ps$_2$ is a complicated nonadiabatic 
four-body system where only the $L=0$ state has been investigated 
so far. The aim of this paper is to explore the spectra of the Ps$_2$ and
the biexciton molecules and to look for their possible bound 
excited states.
\par\indent
To calculate the binding energies of these molecules, the stochastic
variational method \cite{SVM} has been used \cite{prca}. Our trial function is 
assumed in the from:
\begin{equation}
\Phi_{kLS}={\cal A}\lbrace \chi_{SM_S} {\cal Y}_{KLM_L}({\bf v})
{\rm exp} \lbrace -{1\over 2} \sum_{i,j=1}^3
 A_{kij}{\bf x}_i \cdot {\bf x}_j\rbrace\rbrace, 
\end{equation}
with
\begin{equation}
{\bf v}=\sum_{i=1}^{3}u_{ki} {\bf x}_i ,
\end{equation}
where ${\bf x}_1$ and ${\bf x}_2$ are the distance vectors
between the positive
and negative charges in the first and second atom, ${\bf x}_3$ is the
distance vector between the center-of-masses of the two atoms,
$\chi_{SM_S}$
is a spin function, ${\cal Y}_{KLM}({\bf x})=x^{2K+L}Y_{LM}({\hat {\bf x}})$,
``$k$'' is the index of the basis states
and ${\cal A}$ is an antisymmetrizer. The above trial function, the
``correlated gaussian basis'' is known to provide a high quality wave
function and very precise energy \cite{cg,cencek,suv}. 
The correlation between
the particles plays a very important role in the systems
considered and it is efficiently represented by the nondiagonal elements
of $A_k$. The most adequate values of
the nonlinear parameters $u_{ki}$ and $A_{kij}$ are selected 
by a random optimization \cite{prca}. 
The dimension of the basis is increased one by one
until the required convergence is reached. This procedure has proved to be 
efficient, leading to a nearly optimal parameters at a relatively
low computational cost. 
\par\indent
The spectrum of systems made up of positrons and electrons is shown
in Figure I. Both the Ps$^-$ ion and the Ps$_2$ molecule are 
known to have one bound state, and these states have been studied
by quite a few different methods including variational 
\cite{posi,posi1,posi2,Ho,rich} and Quantum Monte Carlo
(QMC) methods \cite{dario1}. As a test, we compare our result
to those of these calculations. The convergence of the energy as a function 
of basis size $N$ is shown in Table I.
Our results for the Ps$_2$ molecule improves the best previous
variational results and it is in a good agreement
with the recent QMC calculation of Bressanini {\it et al.} 
\cite{dario1}.
\par\indent
The main aim of this paper is to seek for other bound excited
states of the Ps$_2$ molecule and biexcitons. To this end we
applied our method for all possible combinations of states
with $L=0,1,2,3$ orbital angular momenta and $S=0,1,2$ spins.
No bound excited
states have been found for $\sigma=1$ except for one case. In the case
of $L=1$ (with negative parity) and $S=0$ our calculation predicts
the existence of a second bound state of the Ps$_2$ molecule.
In that system, the spins of the positronium atoms are coupled to zero.
In this spin state, the Ps$_2$ molecule can dissociate into two
Ps atoms (bosons) only if the relative orbital angular momentum is even.
Consequently, the Ps$_2$ molecule with $L=1$ and negative parity
cannot decay into the ground states of two Ps atoms
(Ps($L=0$)+Ps($L=0$)). The energy of this Ps$_2(L=1)$ state
(E=$-$0.334408 a.u., see Table I.) is lower than the energy of the 
relevant threshold ($-$0.3125 a.u., see Figure I.), and 
this state is therefore stable against autodissociation 
into Ps($L=0$)+Ps($L=1$). The binding energy of this state is  
0.5961 eV, which by about 40\% more than that of the ground state of
Ps$_2$ (0.4355 eV). 
\par\indent
The binding mechanism of this second bound state is very special.
The constituents are fermions, but in the decaying channel they 
form bosons (Ps atoms or excitons). The Pauli principle, however, 
forbids the odd partial
waves between bosons so the biexcitons with $L=1$ and negative parity
cannot decay into two excitons. A somewhat similar situation,
that is a second bound state which cannot decay due to parity
conservation, exists in the H$^-$ ion as well. By changing 
the mass ratio in that $(M+,m-,m-)$ system, however, this 
state disappears, so the Ps$^-$ ion has only one bound state.
\par\indent
Before discussing the spatial distribution of this molecule,
let's recall that the average distance $\langle r_{e^+e^-}\rangle$
between the electron and the positron  is 3 a.u. 
in the ground state of the Ps atom, 
while  it is 10 a.u. in the first excited state. 
The root mean square radius (rms) of the ground state of 
the Ps$_2$ molecule
is found to be 3.614 a.u.. The rms radius of the second bound state
is  5.661 a.u., 1.5 times larger than that of the ground state. 
This is not surprising if one assumes
that the second  bound state is essentially a 
system of a Ps atom in its ground
and a Ps atom in its first (spatially extended) 
excited state. To check the validity of this
assumption we have restricted the model space to include  
only this type of configuration. (This can be achieved by a special choice of 
the $u_{ki}$ parameters in equation (2).)
The energy converged to {$-$}0.323 a.u., that is, the Ps($L=0$)
+Ps($L=1$) system with zero relative orbital angular 
momentum forms a bound state with
energy close to that of $L=1$ state of the Ps$_2$ molecule, therefore
this configuration is likely to be the dominant configuration in
this molecule.
There is a second configuration, the Ps$^-- {\rm e}^+$ 
(or Ps$^+- {\rm e}^-$) with $L=1$ relative 
orbital motion, which intuitively may look important because
two oppositely charged particles attract each other, but 
it is merely bound ($E=-0.315$ a.u).
On the other hand by increasing the mass $M$ of the positively charged
particles toward infinity, one arrives at the energy of the 
$C$ $^1\Pi_u $ $2p\pi$ state of the H$_2$ molecule. This state is formed 
by the excited and the ground states of H atom.  Consequently, 
the second
bound state of the biexciton molecule is dominantly formed 
by an interacting
pair of a ground state exciton and an $L=1$ excited state exciton.
\par\indent
The average distances, the average square distances,
the scale factor $\eta=-V/(2T)$ and the probability of finding two particles
in the same space point are listed in Table II  
for both the $L=0$ and $L=1$ states of the Ps$_2$ molecule. 
The closeness of the  scale factor to unity  
proves the convergence of
the results. The average distances show that in the 
$L=1$ state the two atoms are well separated. One cannot give direct 
geometrical picture of the ground or excited state because the 
variance $\Delta r_{ij}=\sqrt{\langle r_{ij}^2 \rangle 
-\langle r_{ij} \rangle^2}$ is large.
\par\indent
In the positronium  limit $(\sigma=1)$ we deal with antiparticles
and the electron-positron pair can annihilate.
The most dominant annihilation is accompanied by the emission of 
two photons with energy about 0.5 MeV each. To have an estimate for the 
decay due to the annihilation we have substituted the probability 
density of an electron at the position of a positron  
into the formula (64) of Ref. [9]. 
Roughly speaking, the lifetime is inversely proportional to the 
probability of finding an electron and a positron in the same
position ($\langle \delta(r_{e^+e^-})\rangle$, see Table II). 
The lifetime due to the annihilation is estimated to be 1.8 ns. This is
about two times of that of the ground state. 
\par\indent
The dependence of the biexciton binding energy on the mass 
ratio $m/M$ is shown in Figure II. The change of the binding energy 
in the ground and the excited states is similar.
Both the ground and excited states become less bound
by changing the mass ratio from H$_2$ to Ps$_2$, though 
the binding of the
excited state decreases  somewhat to lesser extent.
The energy of the transition from the excited $L=1$ to the 
ground $L=0$ state is
also shown in this figure. This transition may take place in an external 
field, for example. 
\par\indent
The continued advance in microfabrication has allowed 
the creation 
of semiconductor systems  (quantum dots) where the electrons 
(or excitons, biexcitons, etc.) are laterally confined. 
This technical possibility has intensified the interest in 
two-dimensional (2D) systems for the last few years. It is 
intriguing, therefore, to investigate
the existence of the second bound state of the biexcitons in two dimension.
To this end we set the azimuthal angles of all the vectors 
${\bf x}_i$ and ${\bf v}$ to $\pi/2$ and choose $L=M$.
With this particular choice the angular dependence of the wave function
is given by $(v_x+iv_y)^M$ (where $v_x$ and $v_y$ are the $x$ and $y$
component of ${\bf v}$. 
\par\indent
After repeating the calculation for 
the two dimensional  
Ps$_2$ molecule we found that its energy is $-2.192858$ a.u.
This is in good agreement with the 
QMC result  $-2.1928 \pm 0.0001 $\cite{dario2}. 
Note that the energy of the Ps atom in two dimension is
$E_L=-1/(2L+1)^2$. 
The size of this system is 
very small: Its rms radius is 1.125 a.u. which is 
smaller than one third of the radius in 3D. The interesting property 
of these systems in 2D
is that their binding energy considerably increases 
(more than ten times
of that of the 3D case) and their spatial extension is much smaller,
accordingly. 
\par\indent
The energy of the Ps$_2$ molecule with $L=1$ (negative parity) is 
$-$1.2788 a.u. and the rms radius is 2.55 a.u.. By comparing
this energy to the energy of the relevant threshold ($-$1.1111 a.u.)
we can conclude that the second bound state of the Ps$_2$
molecule exists in 2D as well. The binding energy of the second bound state 
in 2D is about 8 times of that of the 3D case. 
\par\indent
A second bound state of the biexcitons which exists 
in both 2D and 3D is reported for the first time.
The existence of this state is due purely 
to the Pauli principle. 
This bound state is in the continuum but its autodissociation 
is forbidden. It is shown that not only 
the ground but the $C$ $^1\Pi_u $ $2p\pi$ excited 
state of the H$_2$-like
molecule also survive the change of mass ratio $\sigma$ of the heavier
and lighter particles in the whole $[0,1]$ interval of $\sigma$.
\par\indent
Recent experiments have shown the existence of the biexciton molecules but
the theoretical and experimental binding energies of biexcitons 
disagree \cite{Miller,singh1,singh2}. 
There are different suggestions to resolve this discrepancy,
e.g., in terms of fractional dimensions, localization, etc. 
All of these models
are based on the fact that the biexcitons under practical conditions
are not pure Coulombic four-body systems in the ``vacuum''. 
Some models assume
that in the biexcitons the Coulomb interaction is slightly modified, 
like e.g. $(1-{\rm exp}(-\gamma r))/r$. Our calculation confirms 
that the second 
bound state survives this distortion of the potential as well. 
Experimental confirmation of the existence of the second bound 
state would be very useful because 
it would help to pinpoint the most realistic
model of the biexcitons in quantum wells. The fact that the binding 
energy is relatively large gives a hope that experimental
measurements for the second bound state is not much more difficult than those
of the ground state.
\bigskip
\par\indent
This work was supported by OTKA grant No. T17298 (Hungary) and 
Grant-in-Aid for International Scientific Research 
(Joint Research) (No. 08044065) of the Ministry of Education, Science 
and Culture (Japan). The authors are 
grateful for the use of RIKEN's computer facility which made possible 
most of the calculations.

\begin{table}

\caption{Total energies of the Ps$_2$ molecule
in atomic unit}

\begin{tabular}{llc}\hline
method  & Ps$_2$($L=0$) & Ps$_2$($L=1$) \\
SVM ($N=100$) & $-$0.515981750         &$-$0.334399869 \\
SVM ($N=200$) & $-$0.516002271         &$-$0.334405027 \\
SVM ($N=400$) & $-$0.516003769         &$-$0.334407971 \\
SVM ($N=800$) & $-$0.516003778         &$-$0.334408112 \\
Ref. [9]      & $-$0.516002            &   \\
QMC  [13]     & $-$0.51601$\pm$0.00001 &   
\end{tabular}
\end{table}

\begin{table}

\caption{Ground and excited state expectation values (in a.u.).}
\begin{tabular}{ccc}\hline
expectation value & Ps$_2$($L=0$) & Ps$_2$($L=1$) \\
\hline
$\langle r_{e^-e^-}^2 \rangle$        & 46.371    & 96.047   \\
$\langle r_{e^+e^-}^2 \rangle$        & 29.111    & 80.152   \\
$\langle r_{e^-e^-}   \rangle$        &  6.033    & 8.856    \\
$\langle r_{e^+e^-}   \rangle$        &  4.487    & 7.568    \\
$\langle \delta(r_{e^-e^-}) \rangle$  &  0.00063  & 0.00015  \\
$\langle \delta(r_{e^+e^-}) \rangle$  &  0.022    & 0.011    \\
$-\langle V \rangle/(2 \langle T\rangle)$ & 0.999999970& 0.9999984
\end{tabular}
\end{table}

\begin{figure}[t]
\centerline{\psfig{figure=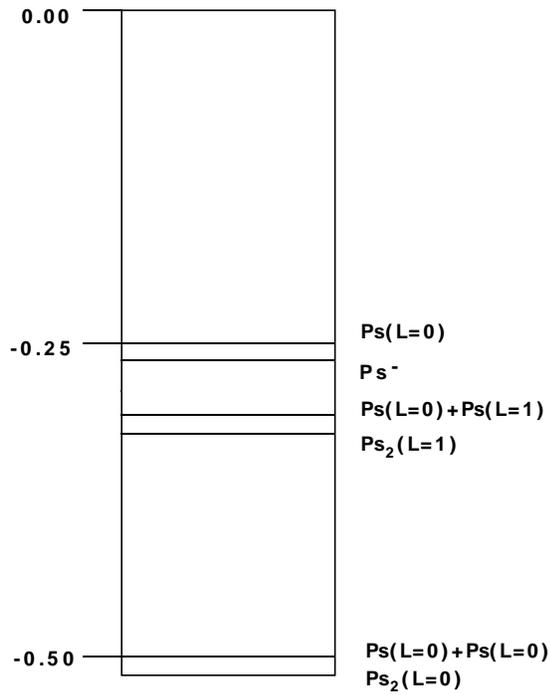}}
\caption{
Bound states of the electron-positron systems.}
\label{fig1}
\end{figure}

\begin{figure}[b]
\centerline{\psfig{figure=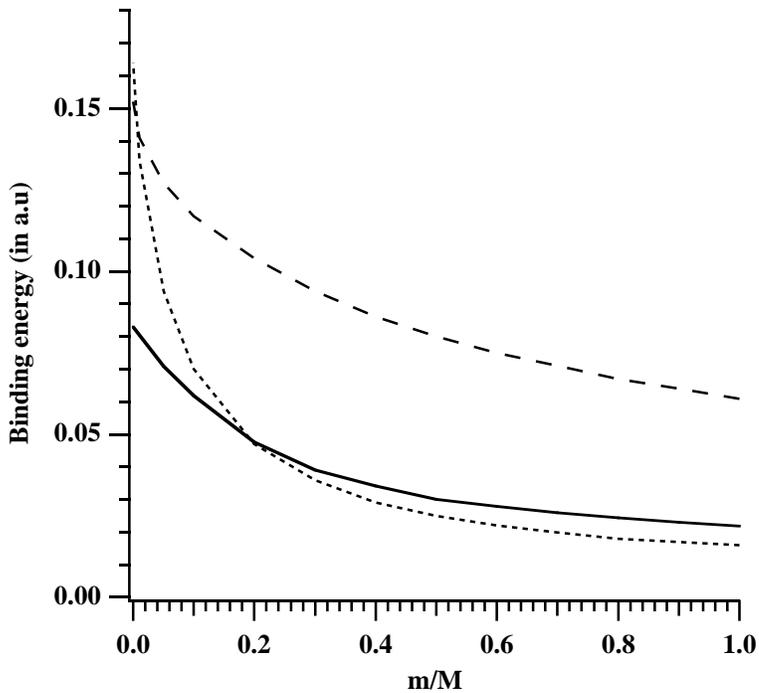}}
\caption{
Energies of the biexcitons as a function of mass ratio. The dotted 
line is the binding energy of the ground state, and the dashed line 
is that of the first excited state. The continuous line shows 
the energy difference of the first excited and the ground states. 
Note that the energy difference is divided by three so as to fit the 
figure.}
\label{fig2}
\end{figure}

\end{document}